\documentclass{article}
\usepackage{spconf,amsmath,graphicx}
\usepackage{booktabs}  
\usepackage{xcolor}
\usepackage{mathtools}
\usepackage{algorithm2e, setspace}
\usepackage{comment}

\title{Efficient Speech Emotion Recognition using Multi-scale CNN and Attention}
\name{Zixuan Peng\sthanks{Equal Contributions.}, Yu Lu\footnotemark[1], Shengfeng Pan, Yunfeng Liu }
\address{Zhuiyi Technology, Shenzhen, China \\ \\
  panacea\_raynor@hotmail.com, \{julianlu, nickpan, glenliu\}@wezhuiyi.com}

\begin{document}
%
\maketitle

\begin{abstract}
Emotion recognition from speech is a challenging task. Recent advances in deep learning have led bi-directional recurrent neural network (Bi-RNN) and attention mechanism as a standard method for speech emotion recognition, extracting and attending multi-modal features - audio and text, and then fusing them for downstream emotion classification tasks. 
In this paper, we propose a simple yet efficient neural network architecture to exploit both acoustic and lexical information from speech. The proposed framework using multi-scale convolutional layers (MSCNN) to obtain both audio and text hidden representations. Then, a statistical pooling unit (SPU) is used to further extract the features in each modality. Besides, an attention module can be built on top of the MSCNN-SPU (audio) and MSCNN (text) to further improve the performance. Extensive experiments show that the proposed model outperforms previous state-of-the-art methods on IEMOCAP dataset with four emotion categories (i.e., angry, happy, sad and neutral) in both weighted accuracy (WA) and unweighted accuracy (UA), with an improvement of 5.0\% and 5.2\% respectively under the ASR setting. \if Moreover, Our model obtains dramatic speedups and has much less memory consumption comparing to previous Bi-RNN based model, which indicates its potential to be applied in practical usage. \fi 

\end{abstract}
%

\section{Introduction}
\label{sec:intro}

Speech-based emotion recognition has raised a lot of attention in both speech and natural language processing in recent years.
Emotion recognition - the task of automatically recognizing the human emotional states (i.e. happy, sad, anger, neutral) expressed in natural speech.
It has been an important sub-task in building an intelligent system in many fields, such as customer support call review and analysis, mental health surveillance,  human-machine interaction, etc. \\
One important challenge in speech emotion recognition is that, very often, the interaction between audio and language can change the expressed emotional states. For example, the utterance `Yes, I did quite a lot' can be ambiguous without knowing prosody information. In contrast, `You know what, I'm sick and tired of listening to you' can be considered neutral if the voice is flat and no lexical content is provided. Thus, it is expected to consider both lexical and acoustic information in emotion recognition from speech.\\
Recently, deep learning based approaches has shown great performance in emotion recognition \cite{Yoon1, Xu2019LearningAlig, Yoon2019MHA}. Recurrent neural networks (RNN) and attention mechanism have demonstrated impressive results in this task. In \cite{Xu2019LearningAlig}, an attention network is used to learn the alignment between speech and text, together with Bi-LSTM network to model the sequence in emotion recognition. In addition, \cite{Yoon2019MHA} proposed a multi-hop attention to select relevant parts of the textual data and then attend to the audio feature for later classification purpose. However, these proposed methods are typically computationally expensive and complex in network structure.\\
In this paper, we first propose a simple convolutional neural network (CNN) and pooling -based model termed as multi-scale CNN with statistical pooling units (MSCNN-SPU), which learns both speech and text modalities in tandem effectively for emotion recognition. Additionally, with an attention module built on top of the MSCNN-SPU, resulting in MSCNN-SPU-ATT, the overall performance can be further improved. In our extensive experiments on the  Interactive  Emotional  Dyadic  Motion  Capture (IEMOCAP) dataset \cite{iemocap}, we show that a) Our MSCNN-SPU model outperforms previous state-of-the-art (SOTA) approaches for bi-modal speech emotion recognition by 4.4\% and 4.3\% relative improvement in terms of WA and UA; b) Attention module (MSCNN-SPU-ATT) can further improve the overall performance by 0.6\% and 0.9\% compare to the MSCNN-SPU.\\
The rest of the work is structured as follows, Section 2 compares our work with prior studies in speech emotion recognition. We then present our proposed model in detail in Section 3. We show the extensive experimental results in Section 4 to compare with previous works, and we conclude the paper in Section 5.

\textbf{Reproducibility.} All our code will be available in open-source on Github\footnote{https://github.com/julianyulu/icassp2021-mscnn-spu}. 

\section{Related Work}
\label{sec:related}
Various approaches to address speech emotion recognition tasks have been investigated using classical machine learning algorithm. For example, previous works studied to model handcrafted temporal features from raw signal using Hidden Markov Models (HMMs) \cite{hmm2003}, or rely on high-level statistical features using Gaussian Mixture Models (GMMs) \cite{gmm2007}.

Benefited from the development of deep learning, many approaches based on deep neural networks (DNNs) have emerged recently. Researchers have demonstrated the effectiveness of CNNs in emotion classification with audio features \cite{audiocnn1, audiocnn2} and text information \cite{textcnn}\if - \cite{ textcnn, charcnn}\fi. Additionally, RNN based models are also investigated to tackle the problem through sequence modeling \cite{textlstm}\if -\cite{textgru, textlstm}\fi. However, either audio or text is used in these methods; while human emotional state is usually expressed through an interaction between speech and text.

Multi-modal approaches make use of both text and audio features. In \cite{multimodal2015}, a hybrid approach using WordNet and part-of-speech tagging are combined with standard audio features, then classified by a Support Vector Machine. Using DNNs, \cite{multimodal2018} extracted text features from multi-resolution CNNs and audio information from BiLSTM, and optimized the task using a weighted sum of classification loss and verification loss. In \cite{Xu2019LearningAlig}, an LSTM network is used to model the sequence for both audio and text. Then, a soft alignment between audio and text is performed using attention. \cite{Yoon1} fused learned features by introducing attention mechanism between text and audio. Specifically, they proposed a so-called multi-hop-attention mechanism to improve the performance and achieved competitive results on the IEMOCAP dataset, which further exploiting the increasingly complicated modelling scheme such as residual learning, attention, etc.

\if However, we do not believe that researchers have properly explored the baseline models in speech recognition, at the time of pushing towards complex model structures, by adding e.g. multi-layer RNNs, attention mechanism over audio and text modality, etc. To this end, our model builds up on simple CNN, pooling and attention, which establishing a strong baseline for future study in Bi-modal emotion recognition from speech. \fi

\if Instead of using RNNs and attention mechanism, we apply only CNNs as the feature extractor, followed by a concatenation layer and softmax layer for classification. Similar to xxxx, the CNN feature extractor are applied at multi-scale with various kernel sizes. However, in our approach the same CNN structure is used for both text and audio signal. Additionally, the global statistic pooling unit (SPU) is used to further enrich the high-level features. Thus, our model is a completely new approach to speech emotion recognition, highlighted with its simple yet effective feature extractor as well as light-weight feature-fusion and classification units. \fi

\section{Model}
\label{sec:model}
In this section,  we will discuss the architecture of our model. We begin with the multi-scale convolutional neural network in use. Next, the grouped parallel pooling layers - named as statistical pooling unit (SPU), and the attention layer are investigated accordingly. 

\begin{figure*}[!htb]

\begin{minipage}[b]{1.0\linewidth}
  \centering
  \centerline{\includegraphics[width=1.0\textwidth]{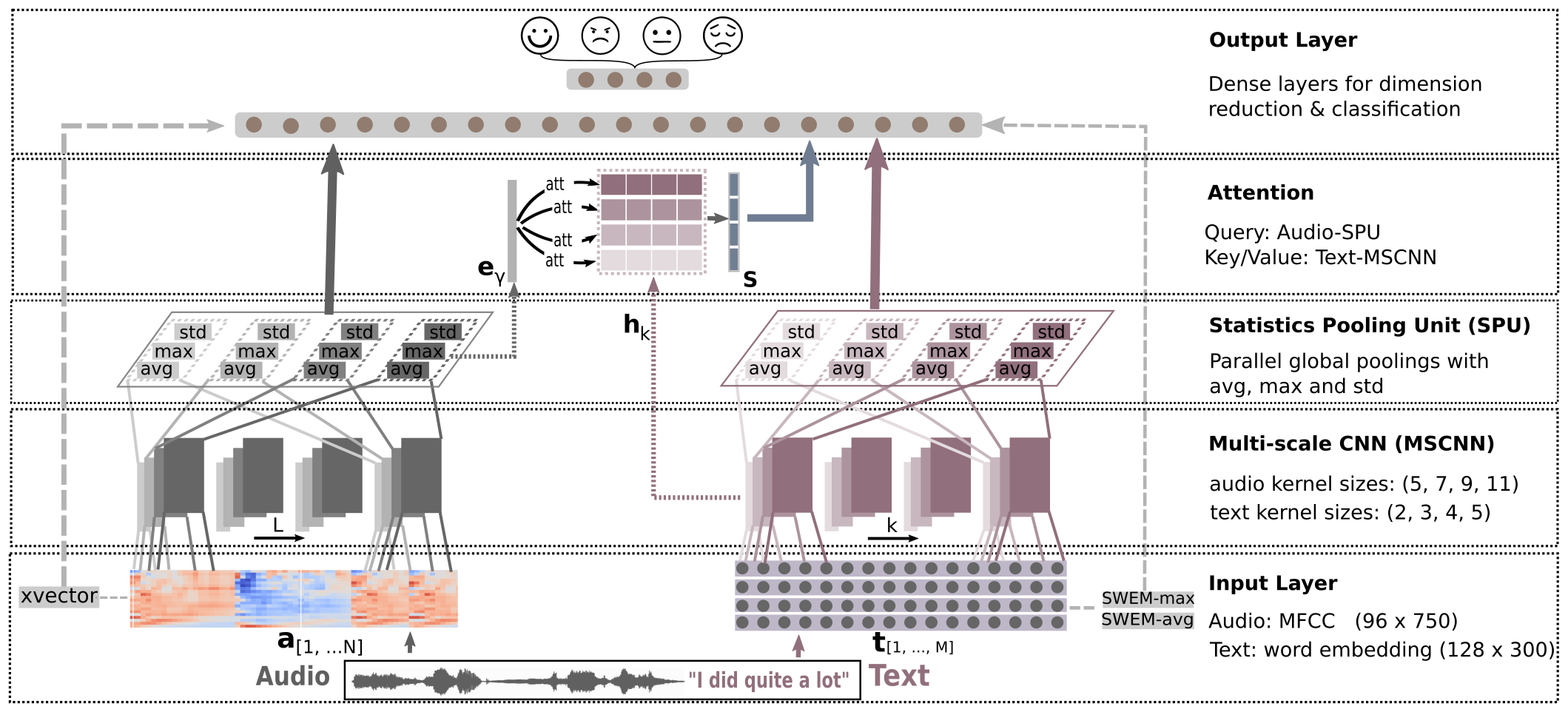}}
\end{minipage}
%
\caption{Architecture of the MSCNN-SPU-ATT model. From bottom to top: a) Input layer: prepare MFCC features from the raw audio and word embedding vectors from the text; b) c) d) MSCNN + SPU + Attention: concatenate features extracted at multiple scales followed by statistic pooling units and attention; e) Output layer: emotion classification with softmax layer after dimension reduction for audio and text feature vectors.}
\label{fig:model}
\end{figure*}

\subsection{Multi-scale CNN}
Motivated by the Text-CNN architecture used in \cite{textcnn}, We adopt and build multiple CNN layers using a group of filters with different kernel sizes for the two separated path, text and audio. We name it as Multi-scale CNN (MSCNN). As shown in Figure \ref{fig:model}, various single layer two-dimensional convolutions with ReLU activation \cite{Nair2010RectifiedLU} are applied in parallel to the input features for text and audio. We employ $[\boldsymbol{a}_1, ..., \boldsymbol{a}_N]$ to represent the sequence of acoustic feature vectors (i.e. Mel-Frequencycepstral Coefficients, or MFCC) in an utterance , where N is the number of frames; $[\boldsymbol{t}_1, ..., \boldsymbol{t}_M]$ represents word embedding vectors, with M indicates the number of tokens in a sentence. Let $\Omega^{A, T} = \{(s, d^{A, T}), s\in S^{A, T}\}$ be sets of kernels for audio modality (superscript `A') and text modality (superscript `T'), where $s$ is the kernel size along the dimension of input sequence, and $d^A$ and $d^T$ are the dimensions of MFCC and word embedding vectors respectively. 

By applying the MSCNN, a set of feature maps are obtained:
\begin{equation}
\text{G}^{\text{MSCNN}}_\alpha(\boldsymbol x, \theta) = \big\{y^{A, T}_\alpha | \alpha \in \Omega\big\}
\end{equation}
\begin{equation}
y^{A, T}_{\alpha = (s, d) \in \Omega}[i,j]  =\sum_{m=\lfloor-\frac{s}{2}\rfloor}^{\lceil\frac{s}{2}\rceil}\sum_{n=\lfloor-\frac{d}{2}\rfloor}^{\lceil\frac{d}{2}\rceil}\boldsymbol{k}[m,n]\cdot \boldsymbol{x} [i-m, j-n]
\end{equation}
where $\boldsymbol{k}$ is the kernel function, and $\boldsymbol{x}$ stands for the input feature. Each of the output feature map $y^{A,T}_{(s, d) \in \Omega}$ is a 2D-matrix which keeps the sequential context while having a series of hidden representations corresponding to the various CNN filters.
\if We experimented with kernel sizes in $\Omega^{A, T}$ and come up with $S^A = \{5, 7, 9, 11\}$ and $S^T = \{3, 5, 7, 9\}$. The Rectified Linear Unit (ReLU)\cite{Nair2010RectifiedLU} is used as activation function throughout the CNNs. 
\fi

\subsection{Statistical Pooling Unit (SPU)}
Previous work with CNNs for emotion recognition task frequently use a single layer of global max-pooling or global average-pooling, and has been proven efficient in \cite{audiocnn1, audiocnn2}. It is intuitive that the abstraction provided by different pooling techniques can help in modeling different emotions. Therefore, we propose a statistic pooling unit (SPU, denoted as $\text{G}^\text{SPU}_{\gamma\in\{\text{max}, \text{avg}, \text{std}\}}$) which consists of three parallel one-dimensional poolings along the sequence modelling direction: a) global max pooling; b) global average pooling; c) global standard deviation pooling, as shown in equation \ref{eq:spu}. \if, where $\Gamma$ is a set of the above pooling operations. \fi  The SPU operation is applied to the output of MSCNN, as shown in the following equation:
\begin{equation}
\label{eq:spu}
\begin{split}
\text{E}= \Big\{\text{G}^{\text{SPU}}_\gamma \big(\text{G}^{\text{MSCNN}}_\alpha(\boldsymbol{x})\big) | \alpha \in \Omega, \gamma \in \{\text{max}, \text{avg}, \text{std}\}  \Big\} \\
\end{split}
\end{equation}

\subsection{Attention}
Inspired by the concept of attention mechanism in \cite{Yoon1, Yoon2019MHA}, we propose a bi-modal attention layer build on top of the audio-MSCNN-SPU and the text-MSCNN. Different from previous work, We consider the outputs from the former as context vectors $\boldsymbol{e}_{\gamma \in \{\text{max}, \text{avg}, \text{std}\}}$ (i.e. the max-pooling, avg-pooling, std-pooling feature 
vectors from the audio branch). The weighting coefficient $s^{\gamma}_{k}$ is computed as a product between the context vectors $\boldsymbol{e}_{\gamma}$ and the $k_{th}$ output feature map from text-MSCNN $\boldsymbol{h}_k$ in terms of the outputs from max-pooling, avg-pooling and std-pooling,  respectively, as shown in Figure \ref{fig:model}. The resulting attention vector $\boldsymbol{S}$ is obtained by weighting $\boldsymbol{h}_k$ with $s^\gamma_k$, as indicated in the following equations:

\if The attention score $s^{\gamma}_{k}$ is used as a weight vector to the text hidden representation $h_k$, to generate the attention vector $S$, as indicated in the following equations:
\fi 
\begin{equation}
    \label{eq:att1}
    s^\gamma_k = \frac{\text{exp}(\boldsymbol{e}^\text{T}_\gamma \boldsymbol{h}_k)}{\sum_k \text{exp}(\boldsymbol{e}^\text{T}_\gamma \boldsymbol{h}_k)} , \text{ where }\gamma \in \{\text{max}, \text{avg}, \text{std}\}
\end{equation}
\begin{equation}
    \label{eq:att2}
    \boldsymbol{S}^\gamma = \sum_k s^\gamma_k \boldsymbol{h}_k
\end{equation}
\begin{equation}
    \label{eq:att3}
    \boldsymbol{S} = \text{concat} (\boldsymbol{S}^{\text{max}}, \boldsymbol{S}^{\text{avg}}, \boldsymbol{S}^{\text{std}})
\end{equation}
\if

The SWEM feature is simply the concatenation of results from global max pooling and global average pooling which directly applied on the word embedding while the embedding matrix is made trainable during back propagation. The resulted SWEM vectors is then concatenated with features extracted from the text-SPU, providing a feature representation from multiple depth.

Xvector is a speaker embedding vector which comes from the hidden layer of TDNN in the pre-trained speaker recognition task\cite{xvector}. The content-independent nature of xvector has been demonstrated to be helpful in the emotion modeling\cite{xvectors-meet-emotions}. 
Similar to SWEM for text, the extracted xvector features are concatenated with the ones from audio-SPU, forming an utterance-level audio feature vector. 
\fi
\subsection{Classification Layer}
For speech emotion classification, audio and text feature vectors from SPU as well as the attended vector are concatenated, combining with SWEM vector which is the concatenation of results from various poolings over the learned word embeddings directly \cite{shen-etal-2018-baseline}. \if and X-vectors \cite{xvector} \fi The resulting vector is passed through a fully-connected layer for dimensionality reduction. Finally, a softmax  layer is used to classify the input 
example into one of the $m $-class emotions, with categorical cross entropy as the loss function:
\begin{equation}
\text{L} = -\sum_{i = 1}^my_ilog(\hat{y}_i) 
\end{equation}


\section{Experiments}
We discuss the dataset, feature extraction, implementation details and evaluation results in this section.
\label{sec:exp}
\subsection{Data}
We use the IEMOCAP dataset \cite{iemocap}, which is a widely used benchmark  in emotion recognition \if SER 
 \fi research. It contains approximately 12 hours of audiovisual data from 10 experienced actors (5 males and 5 females) in both improvised and scripted English conversations. 
For each dialogue, the emotional information is provided in the mode of audio, transcriptions, video, and motion capture recordings. We use audio and transcriptions only in this research. \if There are in total 9 emotional labels and approximately 10000 dialogues. \fi
To be comparable with previous researches \cite{Yoon1,Xu2019LearningAlig, Yoon2019MHA}, 4 categories of emotions are used: \textit{angry} (1103 utterances), \textit{sad} (1084 utterances), \textit{neutral} (1708 utterances) and \textit{happy} (1636 utterances, merged with \textit{excited}), resulting in a total of 5531 utterances. 
Following previous work, we perform a 10-fold cross-validation with 8, 1, 1 in train, dev, test set respectively. Every experiment is run for 3 times to avoid randomness, and the averaged  result is used as the final performance score.

\if On average, each utterances lasts 4.5s (max: 34.1s, min: 0.6s) and contains 11.5 words (min: 1, max: 98). In order to process audio and text examples with CNNs, we set the maximum audio length to 7.5s (mean plus standard deviation) and maximum sentence length to 128. Truncation and padding are used for those that do not meet this standard. \fi

\subsection{Feature Extraction and Implementation Details}
For the audio feature, we use 32-dimensional MFCC feature (frame size 25 ms, hop length 10 ms with Hamming window) combined with its first- and second-order frame-to-frame difference, making it a feature with dimension of 96 in total. The MFCC features are extracted using the librosa \cite{librosa} package. Besides, the X-vector embeddings \cite{xvector} are used as a complementary audio feature, which is extracted from a pre-trained TDNN model on the VoxCeleb dataset \cite{voxceleb} in the speaker identification task, using the Kaldi speech recognition toolkit \cite{kaldi}.
For the text feature, we use 300-dimensional GloVe \cite{pennington2014glove} embedding as the pretrained word embedding for the tokenized transcripts. In addition to the ground-truth text provided by the IEMOCAP database, audio-based ASR transcripts are obtained through the Speech-to-Text API from Google \footnote{Google, Cloud speech-to-text, http://cloud.google.com/speech-to-text/.}. The performance of the Google Speech-to-Text API is evaluated in terms of the word error rate (WER), which yields 5.80\%.

To implement our model, the filter number is set to 128 for every CNN layer. In text encoder, SWEM-max and SWEM-avg features \cite{shen-etal-2018-baseline} are obtained from the word embeddings and then appended to the output of text-SPU. On the other hand, X-vector embeddings are appended to the output of audio-SPU. We minimize the cross-entropy loss using Adam optimizer \if - \cite{Kingma14Adam} \fi with a learning rate of 0.0005. Gradient clipping is employed with a norm 1. The dropout method is applied with a dropout rate of 0.3 for the purpose of regularization. The batch size is set to 64. Besides, The evaluation metrics used are weighted accuracy (WA) \if - each utterance contribute equally in computing accuracy \fi and unweighted accuracy (UA)\footnote{WA: the classification accuracy of all utterances; UA: average of the accuracy from each individual emotion categories}. \if - accuracy is computed within each emotion and then averaged. \fi

\subsection{Performance Evaluation}
\begin{table}
\centering
\caption{Comparison results on the IEMOCAP dataset using speech-only, ground-truth transcript, and ASR processed transcript from Google Cloud Speech API. `A' and `T' represents audio modality and text modality respectively. Bold font indicates best performance.}\smallskip 
\label{tab:main_results}
\smallskip\begin{tabular}[h]{l|ccc}
\toprule
Methods & Modality & WA & UA\\
\midrule
\hline
\multicolumn{4}{c}{Speech-only} \\
\hline
Audio-BRE \cite{Yoon2019MHA} & A & 64.6\% & 65.2\% \\
CNN+LSTM \cite{Satt2017EfficientER} & A & 68.8\% & 59.4\% \\
TDNN+LSTM \cite{Sarma2018EmotionIde} & A & \textbf{70.1}\% & 60.7\% \\
Audio-CNN (ours) & A & 65.4\% & 66.7\% \\
Audio-CNN-xvector (ours) & A & 66.6\% & \textbf{68.4}\% \\
\midrule
\hline
\multicolumn{4}{c}{Ground-truth transcript} \\
\hline
Text-BRE \cite{Yoon2019MHA} & T & 69.8\% & 70.3\% \\
Text-CNN (ours) & T  & 67.8\% & 67.7\% \\
MDRE \cite{Yoon1} & A+T  & 71.8\% & - \\
Learning alignment \cite{Xu2019LearningAlig} & A+T  & 72.5\% & 70.9\% \\
MHA \cite{Yoon2019MHA} & A+T  & 76.5\% & 77.6\% \\
MSCNN-SPU (ours) & A+T & 79.5\% & 80.4\% \\
MSCNN-SPU-ATT (ours) & A+T & \textbf{80.3}\% & \textbf{81.4}\% \\
\midrule
\hline
\multicolumn{4}{c}{ASR transcript} \\
\hline
Text-CNN (ours) & T  & 62.4\% & 61.5\% \\
MDRE \cite{Yoon1} & A+T  & 69.1\% & - \\
Learning alignment \cite{Xu2019LearningAlig} & A+T  & 70.4\% & 69.5\% \\
MHA \cite{Yoon2019MHA} & A+T  & 73.0\% & 73.9\% \\
MSCNN-SPU (ours) & A+T & 77.4\% & 78.2\% \\
MSCNN-SPU-ATT (ours) & A+T & \textbf{78.0}\% & \textbf{79.1}\% \\

\bottomrule
\end{tabular}
\end{table}
\begin{table}
\centering
\caption{Ablation study on proposed model. The gain for each component is shown. }\smallskip 
\label{tab:ab_study}
\smallskip\begin{tabular}[h]{l|ccc}
\toprule
Methods  & WA & UA\\
\midrule
\hline
MSCNN-SPU-ATT  & \textbf{80.3}\% & \textbf{81.4}\% \\
MSCNN-SPU  & 79.5\% & 80.4\% \\
\hline
w/o X-vectors & 78.5\% & 79.3\% \\
w/o Text-SPU (with max-pooling only) & 77.7\% & 78.6\% \\
w/o Text-SWEM & 77.2\% & 78.3\% \\
w/o Audio-SPU (with max-pooling only) & 73.5\% & 74.0\% \\
\bottomrule
\end{tabular}
\end{table}
The experimental results are presented in Table \ref{tab:main_results}. First, we train models with single modality (utterance or ground-truth transcripts only). For speech modality, we use MSCNN+SPU as proposed in Section 3. Besides, we also report the experimental results using Audio-BRE (LSTM) in \cite{Yoon2019MHA}, CNN+LSTM in \cite{Satt2017EfficientER} and TDNN+LSTM in \cite{Sarma2018EmotionIde} for comparison. For text modality, we employ MSCNN+SPU to compare with the Text-BRE \cite{Yoon2019MHA}.
Second, we compare our proposed approach with other multimodal approaches. One straightforward way is to train one LSTM network for each modality, then concatenating the last hidden state from each, as depicted in MDRE \cite{Yoon1}. Learning alignment \cite{Xu2019LearningAlig} employs an LSTM network to model the sequence for both audio and text. Then, a soft alignment between audio and text is performed using attention in the model. In MHA \cite{Yoon2019MHA}, a so-called multi-hop attention is proposed, using hidden representation of one modality as a context vector and apply attention method to the other modality, then repeating such scheme several times. 
As shown in Table \ref{tab:main_results}, Our proposed approach achieves the best results on both WA (80.3\%) and UA (81.4\%) comparing to the other approaches reported in their original papers. 
\if 
Figure \ref{fig:conf_mat} shows the confusion matrix of the test result from our proposed MSCNN-SPU-ATT. the model performs better in angry and sad than the other two emotions; Besides, the model is likely to mis-classify other emotions to neutral the most. 

\begin{figure}[htb]

\begin{minipage}[b]{0.9\linewidth}
  \centering
  \centerline{\includegraphics[width=8.5cm]{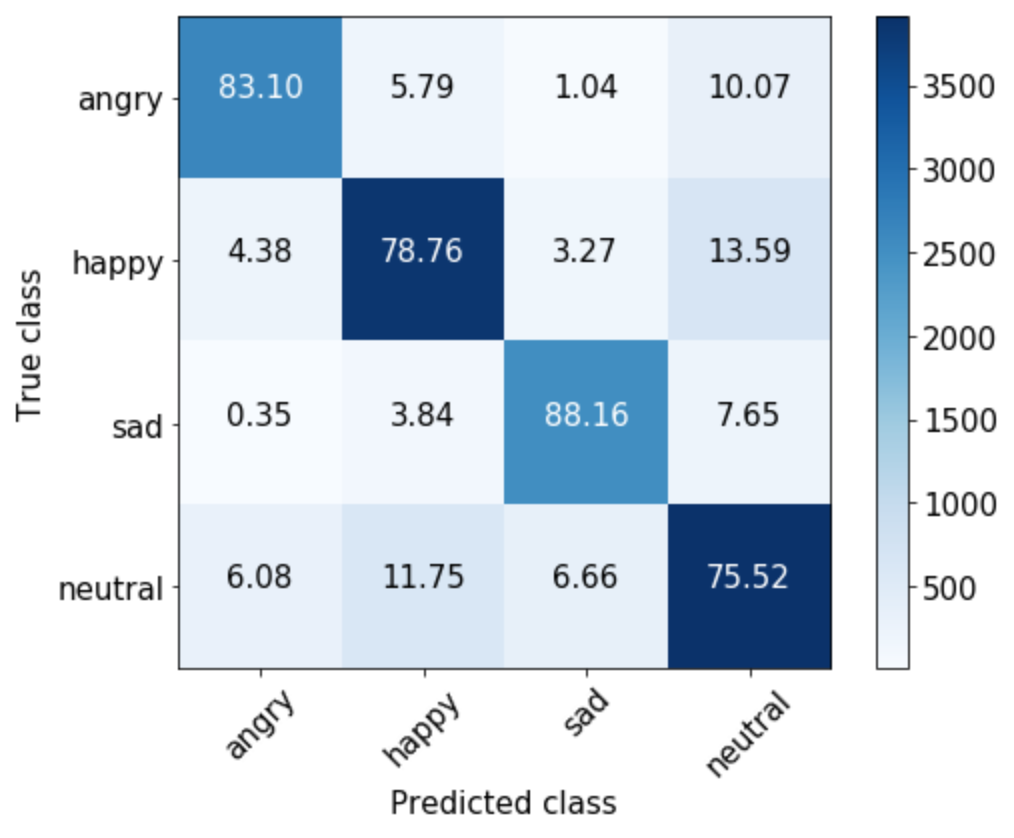}}
\end{minipage}
\caption{Confusion matrix of our proposed model.}
\label{fig:conf_mat}
\end{figure}
\fi
In practical scenario, the ground-truth transcript may not be available. Therefore, we also performed experiments using ASR-processed transcript shown in Table \ref{tab:main_results}. The ASR-processed transcript degrades the performance (roughly 2\%) comparing to ground-truth transcripts. However, the performance of our model is still competitive, specifically, it outperforms the previous SOTA by 5.0\% and 5.2\% in WA and UA respectively. Furthermore,  we conducted an ablation study to analyze the influence of each component in our model, as illustrated in Table \ref{tab:ab_study}.




\section{Conclusions}
\label{sec:conclusions}
In this paper, we proposed a simple yet effective CNN and attention based neural network to solve the emotion recognition task from speech. The proposed model combines audio content and text information, forming a multimodal approach for effective emotion recognition. Extensive experiments show that the proposed MSCNN-SPU-ATT architecture outperforms previous SOTA in 4-class emotion classification  by 5.0\% and 5.2\% in terms of WA and UA respectively in IEMOCAP dataset. The model is further tested on ASR-processed transcripts and achieved competitive results which shows its robustness in real world scenario when ground-truth transcripts are not available. 





\bibliographystyle{IEEEbib}
\bibliography{refs}

\end{document}